\documentclass[twocolumn,showpacs,aps,floatfix,superscriptaddress]{revtex4}
\usepackage{amsmath,amssymb,eucal,graphicx,bm}
\begin{document}
\title{Maxima of Two Random Walks: Universal Statistics of Lead Changes}
%\title{Maxima of Two Random Walks: Universal Statistics of the Number of Lead Changes}
\author{E.~Ben-Naim}
\affiliation{Theoretical Division and Center for Nonlinear Studies,
Los Alamos National Laboratory, Los Alamos, New Mexico 87545, USA}
\author{P.~L.~Krapivsky}
\affiliation{Department of Physics, Boston University, Boston,
Massachusetts 02215, USA}
\affiliation{Institut de Physique Th\'eorique, Universit\'e Paris-Saclay and CNRS, 91191 Gif-sur-Yvette, France}
\author{J. Randon-Furling}
\affiliation{SAMM (EA 4543), Universit\'e Paris-1 Panth\'eon-Sorbonne, 75013 Paris, France}
\begin{abstract}
We investigate statistics of lead changes of the maxima of two
discrete-time random walks in one dimension. We show that the average
number of lead changes grows as $\pi^{-1}\ln t$ in the long-time
limit. We present theoretical and numerical evidence that this
asymptotic behavior is universal. Specifically, this behavior is
independent of the jump distribution: the same asymptotic underlies
standard Brownian motion and symmetric L\'evy flights. We also show
that the probability to have at most $n$ lead changes behaves as
$t^{-1/4}(\ln t)^n$ for Brownian motion and as $t^{-\beta(\mu)} (\ln
t)^n$ for symmetric L\'evy flights with index $\mu$.  The decay
exponent $\beta\equiv \beta(\mu)$ varies continuously with the L\'evy
index $\mu$ for $0<\mu<2$, while $\beta=1/4$ when $\mu>2$.
\end{abstract}
\pacs{05.40.Jc, 05.40.Fb, 02.50.Cw, 02.50.Ey}

\maketitle

\section{Introduction}
\label{sec:Intro}
Extreme values play a crucial role in science, technology, and
engineering. They are linked to rare events, large deviations, and
optimization: minimizing a Lagrangian, for instance, is a frequent
task in physics. Extreme values also provide an important
characterization of random processes, and the study of extreme values
is a significant area in statistics and probability theory
\cite{Gumbel,rse,abn,vbn,jk,mrfky,gw}.

For a scalar random process, there are two extreme values---the
maximum and the minimum. It suffices to consider the maximum. Take the
most basic random process, Brownian motion
\cite{Levy,IM65,BM:book}. By Brownian motion we always mean  
standard Brownian motion, namely the one-dimensional Brownian process
that starts at the origin: $\{B(t): t\geq 0\}$ with
$B(0)=0$.  For the Brownian motion, the maximum process $\{M(t): t\geq
0\}$ is defined by
\begin{equation}
\label{max:def}
M(t)=\max_{0\leq s\leq t} B(s).
\end{equation}

This maximum process is non-trivial as demonstrated by the absence of
stationarity---the increment $M(t+\tau)-M(t)$ depends on both $t$ and
$\tau$. In contrast, the position increment, $B(t+\tau)-B(t)$, depends
only on $\tau$; this stationarity is a crucial simplifying feature of
Brownian motion. Still, the maximum process \eqref{max:def} has a
number of simple properties, e.g. the probability density of the
maximum $m\equiv M(t)$ is a one-sided Gaussian
\begin{equation}
\label{Qmt}
Q(m,t)= \sqrt{\frac{2}{\pi t}}\,\exp\!\left(-\frac{m^2}{2t}\right), 
\quad m\geq 0\,.
\end{equation}
The joint probability distribution of the position and maximum,
$B(t)=x$ and $M(t)=m$, also admits an elegant representation \cite{DC}
\begin{equation}
\label{Pxmt}
\Pi(x,m,t) = \sqrt{\frac{2}{\pi t^3}}
\,(2m-x)\,\exp\!\left[-\frac{(2m-x)^2}{2t}\right]
\end{equation}
This lesser-known formula, which was discovered by L\'evy \cite{Levy},
proves useful in many situations (see \cite{IM65,bk14a}); it can be
derived using the reflection property of Brownian motion.

\begin{figure}[t]
\includegraphics[width=0.44\textwidth]{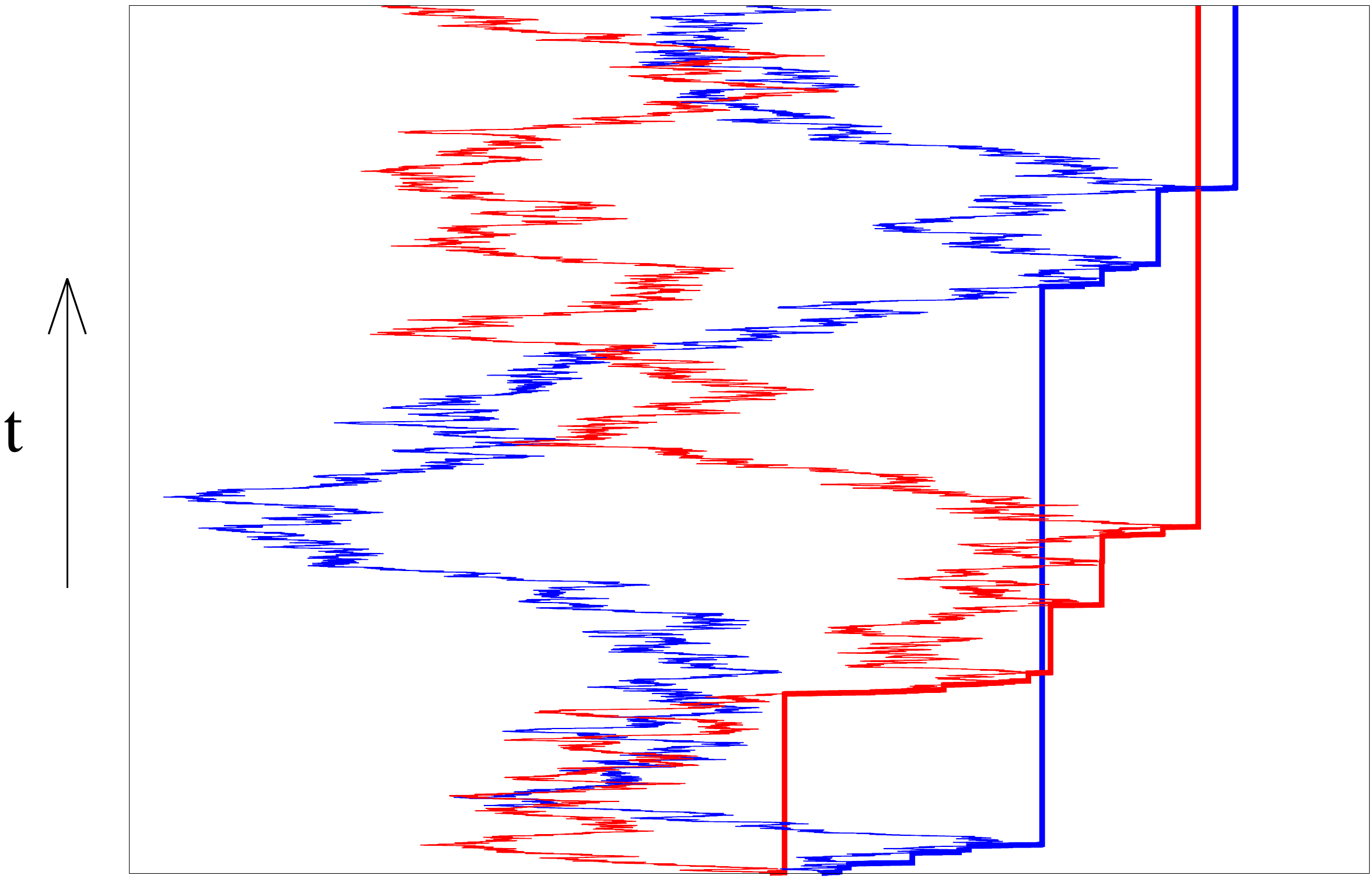}
\caption{Space-time diagram of the positions (thin lines) and the
maxima (thick lines) of two Brownian particles. In this illustration,
the leadership changes twice.}
\label{fig-process}
\end{figure}

In this article, we study the leapfrogging of the maxima of two
Brownian motions (see Fig.~\ref{fig-process}). The probability that
one maximum exceeds another during the time interval $(0,t)$ was
investigated only recently \cite{bk14,JRF}. Here we explore additional
features of the interplay between two Brownian maxima. When
appropriate, we emphasize the universality of our results, e.g. some
of our findings apply to rather general Markov processes, e.g. to
arbitrary symmetric L\'evy flights.

One of our main results is that the average number of lead changes
between the two Brownian maxima exhibits a universal logarithmic
growth
\begin{equation}
\label{nav:log}
\langle n\rangle \simeq \frac{1}{\pi}\,\ln t
\end{equation}
in the long-time limit. This behavior holds for random walks, and more
generally for identical symmetric L\'evy flights.

We also study the probability $f_n(t)$ to have exactly $n$ lead
changes during the time interval $(0,t)$, see
Fig.~\ref{fig-process}. We show that in the case of identical
symmetric L\'evy flights, the probability $f_n(t)$ decays according to
\begin{equation}
 f_n(t) \sim t^{-\beta(\mu)} (\ln t)^n.
 \end{equation}
The persistence exponent $\beta(\mu)$, that characterizes the probability that the lead does not change, 
depends only on the L\'evy index $\mu$. We evaluate numerically the quantity $\beta(\mu)$
for $0< \mu < 2$.

This paper is organized as follows. In Sect.~\ref{sec:ANLC} we
establish the growth law \eqref{nav:log}. First, we provide a
heuristic derivation (Sect.~\ref{sec:HA}) relying on the probability
densities \eqref{Qmt}--\eqref{Pxmt}. We also mention the
generalization to two random walks with different diffusion
coefficients. In Sect.~\ref{sec:FPA} we establish a link with
first-passage time densities and the arcsine law, and we employ it to
provide a more rigorous and more general derivation of \eqref{nav:log}. In
Sect.~\ref{sec:LF} we study the statistics of lead changes for
symmetric L\'evy processes, particularly symmetric L\'evy flights with
index $0<\mu <2 $. The average number of lead changes is shown to
exhibit the same universal leading asymptotic behavior \eqref{nav:log}
independent of the L\'evy index. In Sect.~\ref{sec:Nlc} we show that
for identical symmetric random walks the probability to observe
exactly $n$ lead changes between the maxima behaves as $t^{-1/4}(\ln
t)^n$. We conclude with a discussion (Sect.~\ref{sec:Disc}).

\section{The average number of lead changes}
\label{sec:ANLC}

The question about the distribution of the number of lead changes of
Brownian particles is ill-defined if one considers the
positions---there is either no changes or infinitely many changes. The
situation is different for the maxima: the probability density
$f_n(t)$ to have exactly $n$ changes during the time interval $(0,t)$
is a well-defined quantity if the initial positions
differ. Discretization is still necessary in simulations. The
discrete-time framework also simplifies the heuristic reasoning which
we employ below.  The corresponding results for the continuous-time
case follows from the central-limit theorem: a random walk with a
symmetric jump distribution that has a finite variance converges to a
Brownian motion.

Thus, we consider two discrete time random walks on a one-dimensional
line. The position of each walk evolves according to
\begin{equation}
\label{rw}
x(t+1) = x(t)+\Delta(t)\,.
\end{equation}
The displacements $\Delta(t)$ are drawn independently from the same
probability density $\mathcal{P}(\Delta)$ which is assumed to be
symmetric, and hence $\langle \Delta\rangle =0$. We set the variance
to unity, $\langle \Delta^2\rangle = \int d\Delta\, \Delta^2
\mathcal{P}(\Delta) = 1$. Further, we always assume that the walk
starts at the origin $x(0)=0$. The probability distribution of its
position becomes Gaussian
\begin{equation}
\label{Gauss}
P(x,t) = \frac{1}{\sqrt{2\pi t}}\,\exp\!\left(-\frac{x^2}{2t}\right), 
\end{equation}
in the long-time limit. The maximal position of the walk during the
time interval $(0,t)$ is
\begin{equation} 
\label{mt-def}
m(t)=\text{max}\{x(0),x(1),x(2),\ldots,x(t)\}\,.
\end{equation}
The probability distribution of the maximum is given by \eqref{Qmt}, 
and the joint position-maximum distribution is given by \eqref{Pxmt}.

In the following, we consider two identical random
walkers. We assume that both walkers start at the origin,
$x_1(0)=x_2(0)=0$. Denote by $x_1(t)$ and $x_2(t)$ the positions of
the walkers and by $m_1(t)$ and $m_2(t)$ the corresponding maxima. If
$m_1(t)>m_2(t)$ the first walker is considered to be the leader. If
$m_1(t)=m_2(t)$ the notion of the leader is ambiguous. This happens at
$t=0$, and we postulate that the first walker is the original
leader. There are no lead changes as long as both walkers remain in
the $x<0$ half-line. Once the walkers spend time in the $x>0$
half-line, the identity of the leader can be ambiguous if the
probability density is discrete, e.g.,
\begin{equation}
\label{binary}
\mathcal{P}(\Delta)  = \tfrac{1}{2}\delta(\Delta-1) +  
\tfrac{1}{2}\delta(\Delta+1)
\end{equation}
In the following, we tacitly assume that the probability density does
not contain delta functions, so once the two maxima become positive
they remain distinct.

Eventually the maximum of the second walker will overtake that of the
first and the second walker turns into the leader. Yet, at some later
moment, the leadership will change again. This leapfrogging proceeds
indefinitely. How does the average number of lead changes vary with
time? What is the distribution of the number of lead changes? In this
section we answer the first question. First, we provide a derivation
of the asymptotic growth law \eqref{nav:log} using heuristic
arguments.

\subsection{Heuristic arguments}
\label{sec:HA}

We now argue that the average number of lead changes exhibits the
logarithmic growth \eqref{nav:log}. Let $m_2(t)<m_1(t)$ and the lead
changes soon after that. We use the shorthand notation $m\equiv
m_1(t)$. The probability density for this quantity is given by
\eqref{Qmt}. We are interested in the asymptotic behavior ($t\gg 1$)
and in this regime, the random walker is generally far behind the
maximum. This behavior is intuitively obvious, and it can be made
quantitative---using \eqref{Pxmt} we compute the average size of the
gap between the maximum and the position
\begin{eqnarray}
\label{gap}
\langle m-x\rangle &=& \int_0^\infty dm\int_{-\infty}^m dx\,(m-x)\Pi(x,m,t)  \nonumber \\
&=& \sqrt{\frac{2t}{\pi}}
\end{eqnarray}
The second maximum can overtake the first if $x_2(t)$ is close to
$m$. Since $x_2\leq m_2<m$, both $x_2$ and $m_2$ are close to $m$ and
the probability of this event is 
\begin{equation}
\label{Pmmt}
\Pi(m,m,t) = \sqrt{\frac{2}{\pi t^3}}\,\,m\,\exp\!\left(-\frac{m^2}{2t}\right).
\end{equation}
Once the second particle reaches the leading maximum it will overtake
it given that the first particle is typically far behind, see
\eqref{gap}.  Therefore, the lead changes with rate
\begin{equation*}
\Pi(m,m,t)\, Q(m,t) = \frac{2m}{\pi t^2}\,\exp\!\left(-\frac{m^2}{t}\right)
\end{equation*}
Integrating over all possible $m$ yields 
\begin{equation}
\label{Eqnav}
\frac{d \langle n\rangle}{dt} = \int_0^\infty dm\,\frac{2m}{\pi t^2}\,
\exp\!\left(-\frac{m^2}{t}\right)= \frac{1}{\pi t}
\end{equation}
which gives the announced growth law \eqref{nav:log}.

The average \eqref{nav:log} already demonstrates that the lead
changing process is not Poissonian. For a Poisson process, the
probability $U_n$ to have exactly $n$ lead changes is fully
characterized by the average $\nu(t)\equiv \langle n(t)\rangle$:
\begin{equation}
\label{Un}
U_n=\frac{\nu^n}{n!}e^{-\nu}.
\end{equation}
This distribution would imply that the probability of having no lead
change decays as $\exp[-(\ln t)/\pi]\sim t^{-1/\pi}$, that is, slower
than the $f_0\sim t^{-1/4}$ behavior which has been established
analytically \cite{bk14,JRF}.

It is straightforward to generalize equation \eqref{nav:log} to the
situation where the two random walks have different diffusion
coefficients, denoted by $D_1$ and $D_2$.  By repeating the steps
above, we get
\begin{equation}
\label{Lav_DD}
\langle n\rangle \simeq \frac{2}{\pi}\,\frac{\sqrt{D_1 D_2}}{D_1+D_2}\,\ln t.
\end{equation}
In particular, in the limit when one of the particles diffuses much
more slowly than the other, $\epsilon=D_1/D_2\to 0$, we have 
$\langle n\rangle \simeq (2\sqrt{\epsilon}/\pi)\ln t$. 

\subsection{First-passage analysis}
\label{sec:FPA}

Here, we present an alternative approach which complements the
heuristic arguments given in Sect.~\ref{sec:HA}. This approach is
formulated in continuous time, and it accounts for the universality
observed in the numerical simulations (see Fig.~\ref{fig-nt}).

If $m_1(t)=m > 0$, the probability that $m_2$ becomes the leader in
the interval $(t,t+\Delta t)$ is given by the probability that the
first-passage time of $x_2$ at level $m$ is in $(t,t+\Delta t)$.  Let
us write $\Phi(m,t)$ for the first-passage density, and recall that
the density of the maximum, $Q(m,t)$ is given by \eqref{Qmt}. By
integrating over $m$, we obtain the average rate of lead changes as a
first-passage equation
\begin{equation}
\label{eq-genroc}
\frac{d \langle n\rangle}{dt} = 2 \int_0^\infty dm\, \Phi(m,t) Q(m,t) .
\end{equation}
The multiplicative factor $2$ on the right-hand side of
\eqref{eq-genroc} takes into account that $m_2(t)$ could have been the
leader.  In the case of Brownian motion, the first-passage density
$\Phi(m,t)$ is known \cite{SR_book}, e.g. it can be derived using the
method of images. This allows one to compute the integral in
\eqref{eq-genroc}.

\begin{figure}[t]
\includegraphics[width=0.48\textwidth]{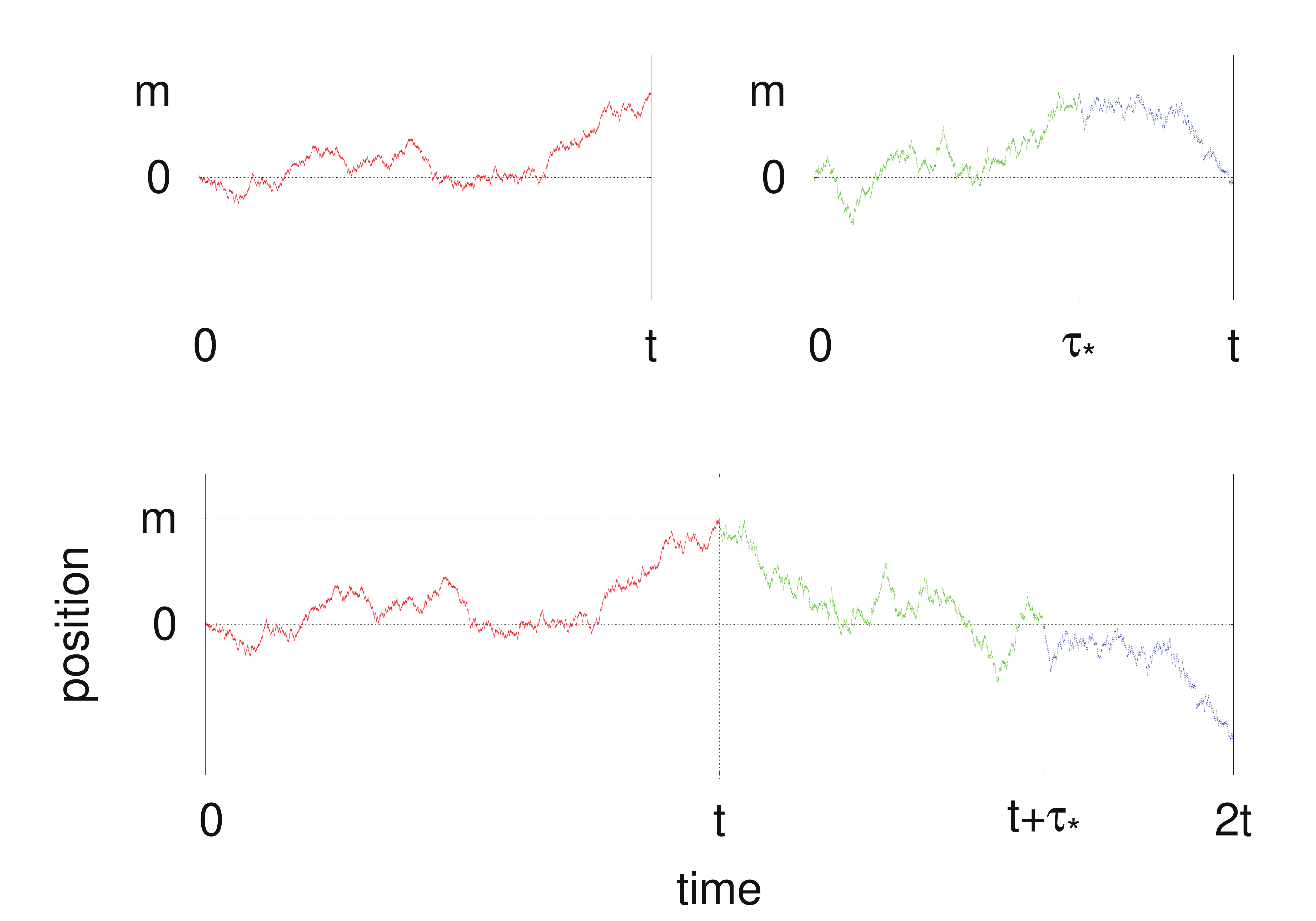}
\caption{Path transformation leading to \eqref{eq-ASL}: split the
second Brownian path (top right) into its pre-maximum part and its
post-maximum part (these two parts are independent by property of the
time at which a Markov process attains its maximum); reverse time in
the pre-maximum part (in green) and shift downward by $-m$ the
post-maximum part (in blue); concatenating this with the other,
independent Brownian path (top left), one obtains a new Brownian path
of double duration (bottom). Note that its final value will be
non-positive, and its maximum, also equal to $m$, will be attained at
time $t$.}
\label{fig-pathtrans}
\end{figure}

Interestingly, one can evaluate the integral in Eq.~\eqref{eq-genroc}
without reference to the explicit forms of $Q$ and $\Phi$. This
evaluation uses path transformations in the same spirit as the
well-known Verwaat construction for Brownian excursions
\cite{Vervaat}. One notices that the integrand in \eqref{eq-genroc}
corresponds to the product of densities associated with two
independent paths: (i) a path first hitting $m$ at time $t$, and (ii)
a path having on $(0,t)$ a maximum equal to $m$ attained at some time
$\tau_*$. One then establishes a bijection between pairs of such paths
and paths of duration $2t$, as illustrated in
Fig.~\ref{fig-pathtrans}.  This path transformation procedure produces paths
of duration $2t$ attaining their maximum at time $t$, and having a \textit{non-positive} final value.

When integrating over $m$ the product $\Phi(m,t) Q(m,t)$, one
therefore obtains one half of the probability density associated with
Brownian paths of duration $2t$ attaining their maximum at time
$t$. This density is easily derived from the famous arcsine law
\cite{Levy2,Levy}, which states that the probability density for the
time $\tau$ at which a Brownian motion attains its maximum over a
fixed time interval $(0,T)$ is given by
$1/[\pi\sqrt{\tau(T-\tau)}]$. One thus obtains 
\begin{equation}
\label{eq-ASL}
\int_0^\infty dm\, Q(m,t) \Phi(m,t)=
\frac{1}{2}\,\frac{1}{\pi \sqrt{t(2t-t)}}=\frac{1}{2\pi t}.
\end{equation}
By combining \eqref{eq-ASL} with \eqref{eq-genroc}, one recovers the
heuristic prediction \eqref{Eqnav} $\left\langle n \right\rangle
\simeq A \ln t$ and confirms $A = 1/\pi$. Moreover, the derivation is
now sufficiently general as it applies to all random walks with
symmetric jump distributions. Hence the leading asymptotic
behavior \eqref{nav:log} does not depend on the details of the
(symmetric) jump distribution, in full agreement with the universality
of the amplitude observed in numerical simulations. The first-passage
equation and path transformation procedure can be applied to
random walks that do not converge to Brownian motion, as discussed in
the next section.

\section{L\'evy flights}
\label{sec:LF}

Brownian motion belongs to a family of (symmetric) L\'evy processes
\cite{Levy,Bertoin,DGNH,LPA}. These processes are stationary,
homogeneous and stable. In a discrete time realization, a L\'evy
process becomes a L\'evy flight. L\'evy flights are ubiquitous in
Nature, see \cite{Man82,BG90,BD97,SM}. Here, we examine lead changes of the maxima of two identical
L\'evy flights. For L\'evy flights, the jump distribution
$\mathcal{P}(\Delta)$ has a broad tail, and the L\'evy index $\mu$
quantifies this tail, $\mathcal{P}(\Delta)\sim |\Delta|^{-\mu -1}$ as
$|\Delta|\to\infty$. For simplicity, we choose the purely algebraic
jump distribution,
\begin{equation}
\mathcal{P}(\Delta) = 
\begin{cases}
\frac{\mu}{2}|\Delta|^{-\mu-1} & |\Delta|>1\,,\\
0                                             & |\Delta|<1\,.
\end{cases}
\end{equation}
When $\mu>2$, L\'evy flights are equivalent to an ordinary random
walk.  For true L\'evy flights $0<\mu\leq 2$; in this range the
variance $\langle \Delta^2\rangle$ diverges. For 
$0<\mu\leq 1$ even the average jump length $\langle
|\Delta|\rangle$ becomes infinite. Generally for $\mu<2$ the displacement $x(t)$ scales as $t^{1/\mu}$.

We now show that the average number of lead changes exhibits
logarithmic growth. Denote by $Q_\mu(m,t)$ the probability density of
the maximum process and by $\Phi_\mu(m,t)$ the density of the first
passage time at level $m$. Following the same reasoning as in
Sect.~\ref{sec:FPA}, we write for the rate at which lead changes take
place:
\begin{equation}
\label{eq-Levyroc}
\frac{d \langle n\rangle}{dt} = 2 \int_0^\infty dm\,\Phi_\mu(m,t) Q_\mu(m,t).
\end{equation}
We now invoke the scaling properties of L\'evy flights
\begin{equation}
\label{laws}
\begin{split}
\Phi_\mu(m,t) &\simeq   m^{-\mu}\Phi_\mu(1,m^{-\mu}t),\\
Q_\mu(m,t)    &\simeq   t^{-1/\mu}Q_\mu(t^{-1/\mu}m,1).
\end{split}
\end{equation}
These scaling forms reflect that the maximum process is $(\mu)$-stable
and the first-passage time process is $(1/\mu)$-stable. Combining
\eqref{eq-Levyroc} and \eqref{laws} we get
\begin{eqnarray*}
\frac{d \langle n\rangle}{dt} &=& 2 \int_0^\infty dm\,m^{-\mu}\Phi_\mu(1,m^{-\mu}t)t^{-1/\mu}Q_\mu(t^{-1/\mu}m,1)\\
&\simeq & \frac{2}{t} \int_0^\infty \frac{dz}{z^\mu}\, \Phi_\mu(1,z^{-\mu})Q_\mu(z,1) = \frac{A(\mu)}{t}.
\end{eqnarray*}
Consequently, the average number of lead changes grows logarithmically
with time,
\begin{equation}
\label{nav:Levy}
\langle n\rangle \simeq A(\mu)\,\ln t.
\end{equation}

\begin{figure}[t]
\includegraphics[width=0.42\textwidth]{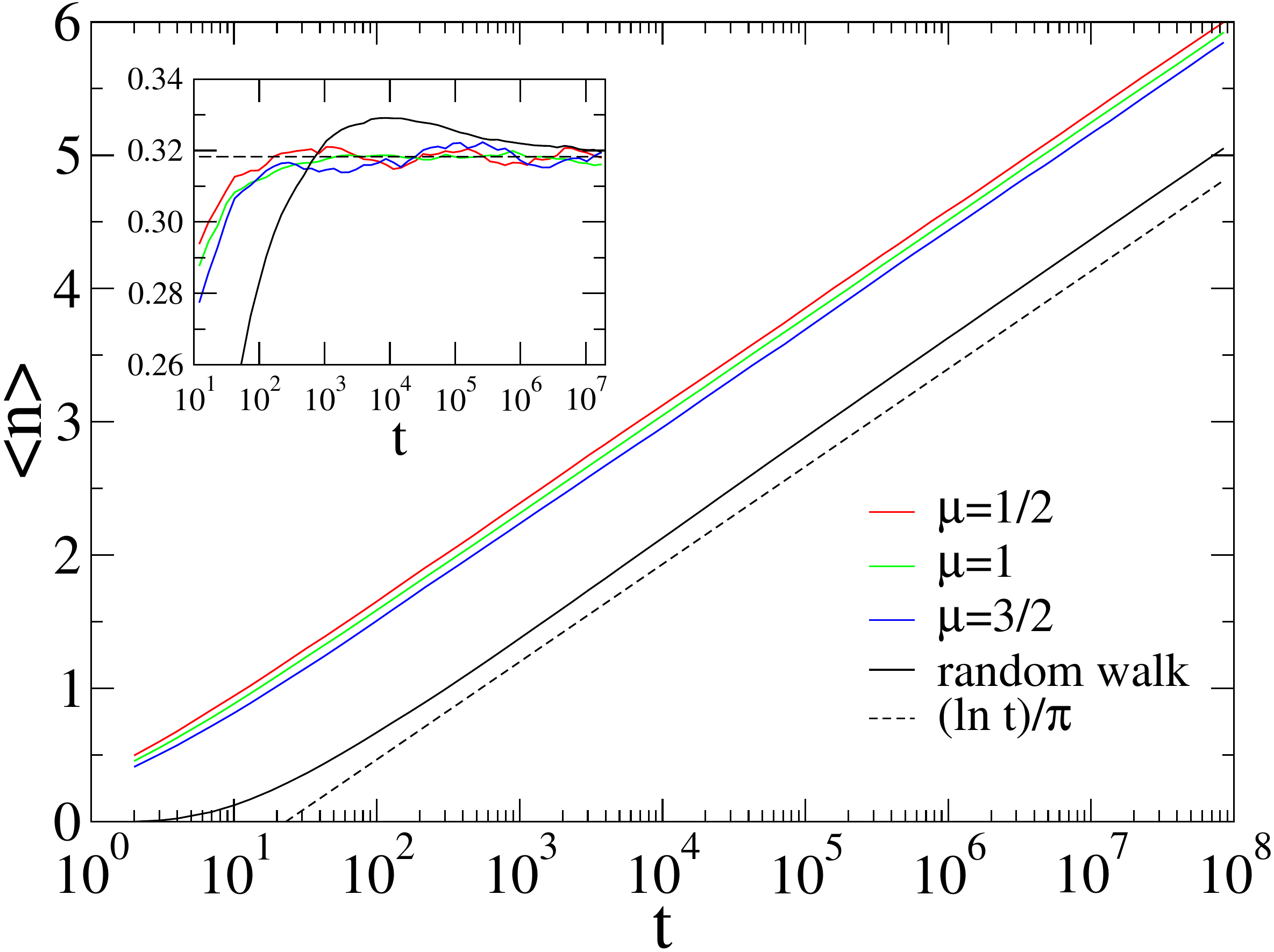}
\caption{The average number of lead changes $\langle n\rangle$ versus
time $t$. Shown (top to bottom) are simulation results for identical
L\'evy flights with index $\mu=1/2, \mu=1, \mu=3/2$ and for random
walks ($\mu>2$).  The inset shows $d\ln\langle n\rangle/d\ln t$ versus
time $t$, demonstrating that the amplitude $A(\mu)$ in
\eqref{nav:Levy} is universal and equal to $1/\pi$.}
\label{fig-nt}
\end{figure}

Our numerical simulations confirm that for identical random walks on a
line, and even for random walks on the lattice with the discrete jump
distribution \eqref{binary}, the amplitude $A(\mu)$ is independent of 
the details of the jump distribution. Furthermore, the universality
continues to hold for the L\'evy flights, so $A=1/\pi$ independent of
the L\'evy index $\mu$. This is confirmed by numerical simulations (see Fig.~\ref{fig-nt}).

\begin{figure}[t]
\includegraphics[width=0.44\textwidth]{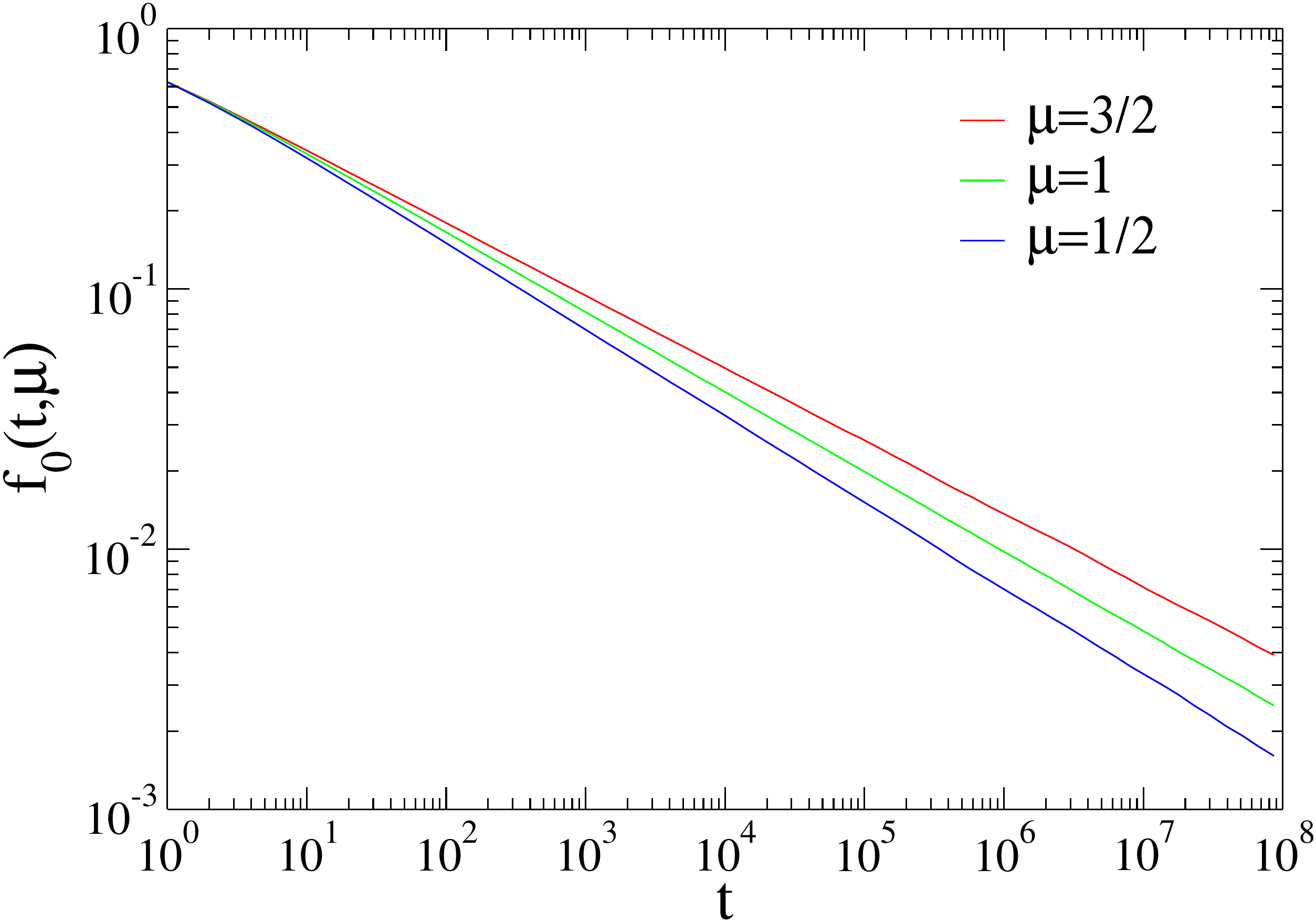}
\caption{The survival probability $f_0(t,\mu)$ versus time $t$ for
L\'evy flights. Shown are (from top to bottom) simulation results for
the values $\mu=3/2$, $\mu=1$, and $\mu=1/2$.}
\label{fig-f0}
\end{figure}

The universality can be understood by noting that the path
transformation procedure described at the end of the previous section,
as well as the arcsine law, are valid for L\'evy flights. The key
requirements are to have cyclic exchangeability of the jumps (which is
guaranteed when these are independent and identically distributed), a
continuous and symmetric jump distribution, and a certain type of
regularity for the supremum, see \cite{F68b,CUB,Bertoin,LPA}. The
validity of the arcsine law implies that we can use
\eqref{eq-ASL}. Thus the leading asymptotic \eqref{nav:log} holds for
a wide class of symmetric L\'evy processes, and in particular,
symmetric L\'evy flights.

\begin{figure}[t]
\includegraphics[width=0.44\textwidth]{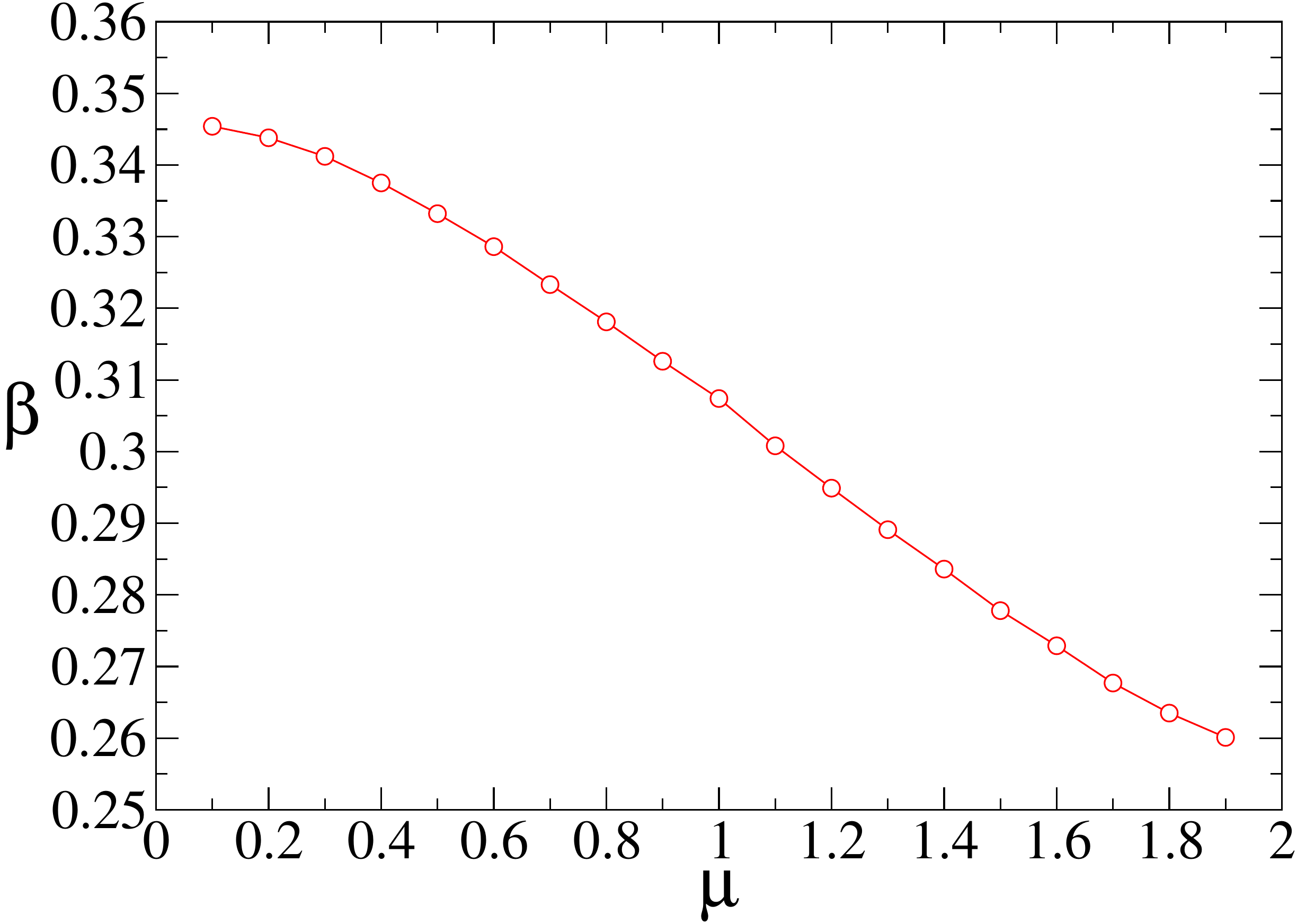}
\caption{The persistence exponent $\beta$ versus the index $\mu$
characterizing (symmetric) L\'evy flights.}
\label{fig-bm}
\end{figure}

Next, we examine the probability $f_0(t,\mu)$ that there are no lead
changes till time $t$. Our simulations show that this probability
decays algebraically (see Fig.~\ref{fig-f0}):
\begin{equation}
\label{f_0}
f_0(t) \sim t^{-\beta(\mu)}. 
\end{equation}
The persistence exponent $\beta(\mu)$ varies
continuously with $\mu$, namely $\beta(\mu)$ is a monotonically
decreasing function of $\mu$, see Fig.~\ref{fig-bm}. For $\mu>2$, 
L\'evy flights are equivalent to ordinary random walks, so
$\beta(\mu)=\frac{1}{4}$ for $\mu>2$.

In the marginal case $\mu=2$, the mean-square displacement has a
logarithmic enhancement over the classic diffusive growth, $\langle
x^2\rangle \sim t\ln t$. Consequently, the convergence toward the
ultimate asymptotic behavior is very slow near $\mu\approx 2$.  Our
numerical simulations suggest that in the marginal case $\mu=2$, there
may be a logarithmic correction to the algebraic decay \eqref{f_0}.

\section{Distribution of the number of lead changes}
\label{sec:Nlc}

We now focus on the probability $f_n(t)$ to have exactly $n$ lead
changes until time $t$. For two identical random walks, the
probability that there are no lead changes decays as $f_0 \sim t^{-1/4}$ in
the long-time limit \cite{bk14,JRF}. Since $f_n(t)$ is the probability that the $(n+1)$-st lead change takes place after
time $t$, the probability density for the $(n+1)$-st lead change to
occur at time $t$ is given by $-df_n/dt$. In particular, from $f_0$,
one can write down the conditional probability density $\rho_s$ of the
time of the next lead change, given that it takes place after time
$s$:
\begin{equation}
\label{eq-nlc}
\rho_s(\tau)=\frac{1}{f_0(s)}\left[ -\frac{df_0(\tau)}{d\tau}\right]\sim s^{1/4}\tau ^{-5/4}.
\end{equation}

Furthermore, the quantity $f_1(t)$ is the probability that the first
lead change occurs at some time $s \leq t$ and that the next one
occurs at some time $\tau \geq t$. The probability density associated
with such a configuration is simply proportional to
\begin{equation*}
\left[-\frac{df_0}{ds}\right] \times \int_t^{\infty}\, d\tau\, \rho_s(\tau) 
\sim s ^{-5/4}\times  s^{1/4} t ^{-1/4} \sim s^{-1}t^{-1/4}.
\end{equation*}
Integrating over $s$ [we could introduce a cut-off $\Delta s$, but
this is akin to a discretization of time, so in the large $t$ limit we
might as well set $1< s < t-1$], we find
\begin{equation*}
f_1(t) \sim \int_{1}^{t-1} ds\, s^{-1}t^{-1/4} \sim t^{-1/4}(\ln t).
\end{equation*}

\begin{figure}[t]
\includegraphics[width=0.44\textwidth]{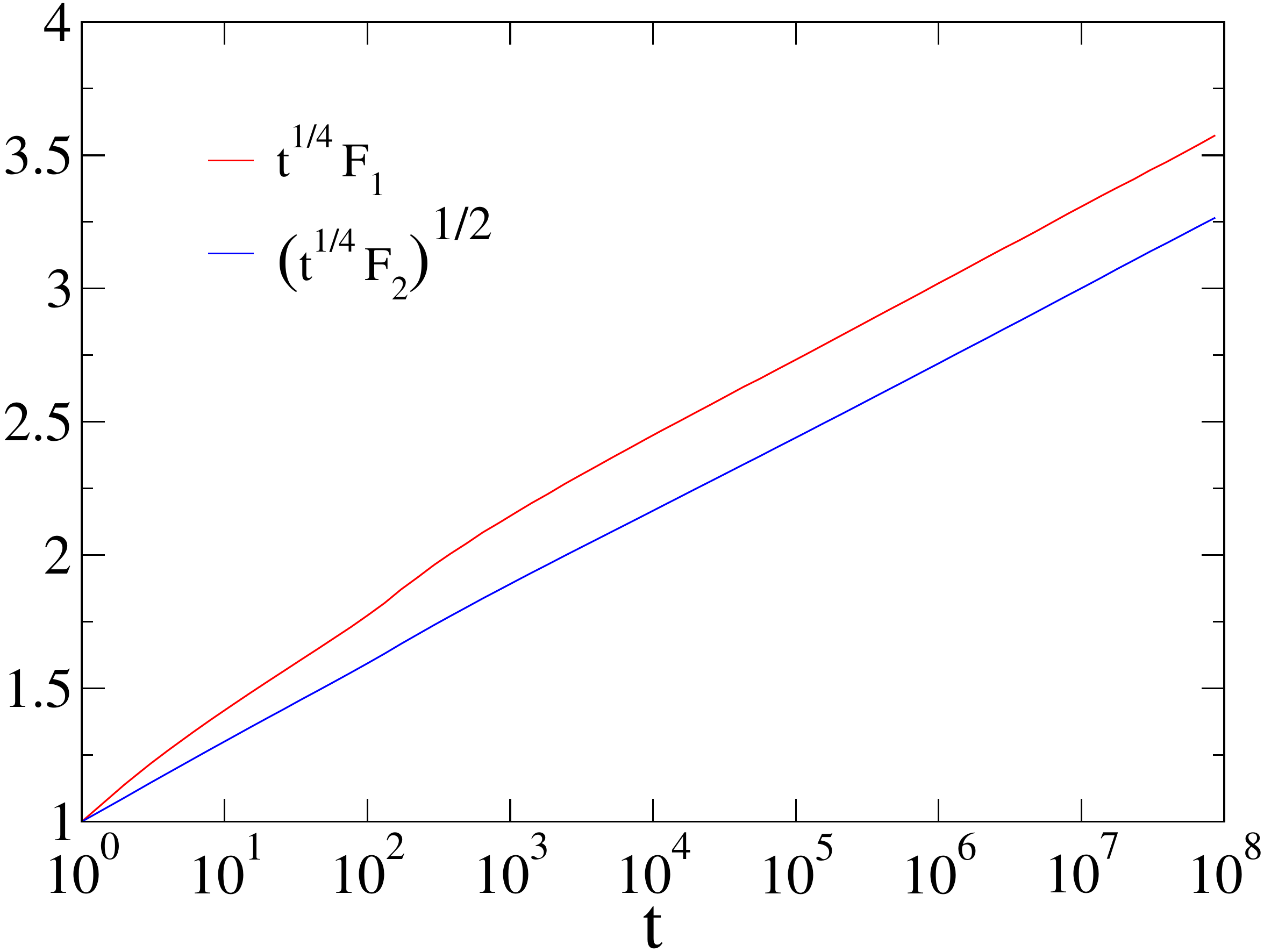}
\caption{The quantities $(F_nt^{1/4})^{1/n}$ versus time $t$ for $n=1$
and $n=2$. The data corresponds to simulations of two ordinary random
walks in one dimension.}
\label{fig-fn}
\end{figure}

Hence, there is a logarithmic enhancement of the probability to have
one lead change compared with having none. The above argument can be
generalized to arbitrary $n$ to yield
\begin{equation}
\label{fn}
f_n\sim t^{-1/4} (\ln t)^n.
\end{equation}
One can establish this general behavior by induction:
\begin{eqnarray*}
f_{n+1}(t)&\sim& \int_1^{t-1} ds\, \left[-\frac{df_n(s)}{ds}\right] \int_t^{\infty}\, d\tau\, \rho_s(\tau) \\
      &\sim& \int_1^{t-1} ds\, (\ln s)^n\,s^{-5/4} s^{1/4}t^{-1/4} \\
      &\sim& t^{-1/4}\int_{1}^{t-1} ds (\ln s)^n\,s^{-1}\\
      &\sim& t^{-1/4}(\ln t)^{n+1}.
\end{eqnarray*}
Thus, there are logarithmic corrections for all $n$.  

We probed the behavior of the cumulative distribution
$F_n(t)=\sum_{0\leq k\leq n}f_k(t)$ using numerical simulations. The
quantity $F_n(t)$ is the probability that the number of lead changes
in time interval $(0,t)$ does not exceed $n$. The dominant
contribution to $F_n$ is provided by $f_n$ and hence
\begin{equation}
\label{Fn}
F_n(t)\sim t^{-1/4}(\ln t)^n. 
\end{equation}
We verified Eq.~\eqref{Fn} numerically for $n=1$ and $n=2$,
see Fig.~\ref{fig-fn}.

For L\'evy flights, the distribution of the number of lead changes can be established in a similar manner.  Using \eqref{f_0} and following the derivation of \eqref{fn} we get
\begin{equation}
\label{fn:mu}
f_n(t) \sim t^{-\beta(\mu)} (\ln t)^n,
\end{equation}
where $\beta(\mu)$ is the aforementioned persistence exponent. Using numerical simulations, we verified that \eqref{fn:mu} holds for $n=1$ and $n=2$ for a few representative values of the L\'evy index in the $\mu<2$ range.

\section{Discussion}
\label{sec:Disc}

The simplest model of correlated random variables is a one-dimensional
discrete-time random walk. Its maximum evolves by a random process
that exhibits a number of remarkable features. Some of these
properties are classical \cite{Levy,IM65,BM:book}, while others were
discovered only recently \cite{bk14a,MZ08,MMS13,GMS14}.  For instance,
the probability distribution for the total number of distinct maximal
values (records) achieved by the walk is universal \cite{MZ08}, i.e.,
independent of the details of the jump distribution, as long as the
jump distribution is symmetric and not discrete. This universality
also holds for symmetric L\'evy flights, and in fact, it is rooted in
the Sparre Andersen theorem \cite{SA54}. In the case of multiple
identical random walks, the total number of records has been
investigated in \cite{WMS}.

In this article, we studied a sort of ``competition'' between maxima
of two identical random walks. In particular, we examined the average
number of lead changes and the probability $f_n(t)$ to have exactly
$n$ changes. We found that the average number of lead changes exhibits
a universal logarithmic growth \eqref{nav:log}. This asymptotic
behavior also holds for symmetric L\'evy flights with arbitrary
index. In contrast, the probability distribution $f_n(t)$ is not
universal: the persistence exponent $\beta(\mu)$ in \eqref{fn:mu}
depends on the L\'evy index $\mu$. The most interesting challenge for
future work is to determine analytically the continuously varying
exponent $\beta(\mu)$ in the range $\mu<2$. 

One natural generalization of the competing maxima problem is to an
arbitrary number $k$ of identical random walks or L\'evy flights. We
expect that the average number of lead changes among the maxima of $k$
identical random walkers still exhibits a universal logarithmic
growth, albeit with a $k$-dependent prefactor. Another quantity of
interest is the survival probability $S_k(t, \mu)$ that $k$ L\'evy
maxima remain ordered, $m_1(\tau)>m_2(\tau)>\ldots > m_k(\tau)$, 
for $\tau=1,2,\ldots,t$. Based on \eqref{f_0}, we anticipate that the
probability $S_k(t, \mu)$ decays algebraically, 
\begin{equation}
\label{Skb}
S_k(t, \mu) \sim t^{-\beta_k(\mu)}. 
\end{equation}
Similar algebraic decays with $k$-dependent exponents describe the probability that the positions of random walks remain perfectly ordered, the probability that one random walk remains the leader, etc. \cite{KM59,mef,BG,G99,bjmkr,bk-mult}.  

For $k=2$, the exponent appearing in \eqref{Skb} is our basic persistence exponent: 
$\beta_2(\mu)\equiv \beta(\mu)$. The exponents $\beta_k(\mu)$ do not
depend on $\mu$ when $\mu>2$, i.e., for random walks. For ordinary
random walks, the exponents $\beta_k$ were studied numerically in
\cite{bk14}, while approximate values for these exponents were
computed in \cite{bkl}.

Generally one would like to explore the statistics of ordering and lead changes for a collection of random variables. Even the case of Markovian random variables is far from understood. Non-Markovian random variables appear intractable, but sometimes the progress is feasible due to hidden connection to Markovian processes. For instance, in the case of Brownian maxima the first-passage processes are Markovian. This allows one to use results from classical fluctuation theory to derive the persistence exponent $\beta=1/4$, see \cite{JRF}, and yields another derivation of the logarithmic growth of the average number of lead changes \cite{CKK}. For L\'evy flights, however, the existence of leapovers at the first passage time of certain points \cite{KLCKM} leads to a breakdown of this approach. New results in this direction represent challenges for future work. 

%\bigskip
\section*{Acknowledgments}
We benefited from discussions with S.~N.~Majumdar. Two of us (PLK and
JRF) thank the Galileo Galilei Institute for Theoretical Physics for
hospitality during the program on ``Statistical Mechanics,
Integrability and Combinatorics'' and the INFN for partial
support. The work of EBN was supported through US-DOE grant
DE-AC52-06NA25396.


\begin{thebibliography}{99}

\bibitem{Gumbel}
       E.~I.~Gumbel,   
       {\it Statistics of Extremes} 
       (Dover, New York 2004).

\bibitem{rse}
       R.~S.~Ellis, 
       {\it Entropy, Large Deviations, and Statistical Mechanics}
       (Springer, Berlin 2005).

\bibitem{abn}
       B.~C.~Arnold, N.~Balakrishnan and H.~N.~Nagraja, 
       {\it Records} (Wiley-Interscience, 1998). 
     
\bibitem{vbn}
       V.~B.~Nevzorov, {\em Records: Mathematical Theory},  
       Translation of Mathematical Monographs {\bf 194} 
       (American Mathematical Society, Providence, RI, 2001).

\bibitem{jk}
      J.~Krug,
      J. Stat. Mech. P07001 (2007).
      
\bibitem{mrfky}
      S.~N.~Majumdar, J. Randon-Furling, M.~J.~Kearney, and M.~Yor, J. Phys. A {\bf 41}, 365005 (2008).

\bibitem{gw}
      G.~Wergen,
      J. Phys. A {\bf 46}, 223001 (2013).
       
\bibitem{Levy}
      P.~L\'evy, 
      {\it Processus Stochastiques et Mouvement Brownien} 
      (Gauthier-Villars, Paris, 1948).

\bibitem{IM65} 
      K.~It\^{o} and H.~P.~McKean, 
      {\it Diffusion Processes and Their Sample Paths} 
      (Springer, New York, 1965).
      
\bibitem{BM:book}
      P. M\"{o}rters and Y. Peres, {\it Brownian Motion}
      (Cambridge: Cambridge University Press, 2010).       

\bibitem{DC}
      In Eqs.~\eqref{Qmt}--\eqref{Pxmt}, we set the diffusion coefficient to $D=\frac{1}{2}$; 
      the general case of an arbitrary diffusion coefficient is recovered by the transformation $t\to 2Dt$.

\bibitem{bk14a}  
      E.~Ben-Naim and P.~L.~Krapivsky, 
      J. Phys. A {\bf 47}, 255002 (2014).

\bibitem{bk14}
      E.~Ben-Naim and P.~L.~Krapivsky,
      Phys. Rev. Lett. {\bf 113}, 030604 (2014).

\bibitem{JRF} 
      J.~Randon-Furling,
      EPL {\bf 109}, 40015 (2015).

\bibitem{SR_book} 
      S.~Redner, 
      {\it A Guide to First-Passage Processes} 
      (Cambridge University Press, Cambridge, 2001).

\bibitem{Vervaat}
      W. Vervaat, Ann. Probab. {\bf 7}, 143 (1979).

\bibitem{Levy2}
      P.~L\'evy, Compositio Mathematica {\bf 7}, 283 (1940).
      
\bibitem{Bertoin} 
      J. Bertoin, 
      {\it L\'evy Processes} 
      (Cambridge University Press, Cambridge, 1996).

\bibitem{DGNH}
       B. Dybiec, E. Gudowska-Nowak, and P. H\"anggi,
       Phys. Rev. E {\bf 73},  046104 (2006).
  
\bibitem{LPA}
       A. E. Kyprianou, 
       {\it Fluctuations of L\'evy Processes with Applications}
      (Springer-Verlag, New York, 2014). 

\bibitem{Man82} 
      B.~Mandelbrot, 
      {\it The Fractal Geometry of Nature} 
      (W. H. Freeman, New York, 1982).

\bibitem{BG90} 
      J.-P. Bouchaud and A. Georges,  Phys.\ Rept.\ {\bf 195}, 127 (1990).

\bibitem{BD97}  
     B.~Derrida, Physica D {\bf 107}, 186 (1997). 

\bibitem{SM}
      S.~N.~Majumdar, Physica A {\bf 389}, 4299 (2010).
      
\bibitem{F68b} 
      W.~Feller, {\it An Introduction to Probability Theory and its Applications}, 
      Vol.~II (Wiley, New York, 1968).

\bibitem{CUB}
       L. Chaumont and G. Uribe Bravo, in: 
       {\it XI Symposium on Probability and Stochastic Processes} (Springer International Publishing, 2015).

\bibitem{MZ08}
      S.~N.~Majumdar and R. M. Ziff, Phys. Rev. Lett. {\bf 101}, 050601 (2008).

\bibitem{MMS13}
      S.~N.~Majumdar, P. Mounaix, and G.~Schehr, Phys. Rev. Lett. {\bf 111}, 070601 (2013).
      
\bibitem{GMS14}
      C. Godr\`eche, S.~N.~Majumdar, and G.~Schehr, J. Phys. A {\bf 47}, 255001 (2014).
      
\bibitem{SA54}
       E. Sparre Andersen, Math. Scand. {\bf 1}, 263 (1953); {\it ibid}. {\bf 2}, 195 (1954). 

\bibitem{WMS}
      G. Wergen, S.~N.~Majumdar, and G. Schehr, Phys. Rev. E {\bf 86}, 011119 (2012).

\bibitem{KM59}
      S. P. Karlin and G. MacGregor,  Pacific. J. Math. {\bf 9}, 1141 (1959).

\bibitem{mef}
      M.~E.~Fisher,
      J.~Stat. Phys. {\bf 34}, 667 (1984).

\bibitem{BG} 
      M. Bramson and D. Griffeath,
      in: {\it Random Walks, Brownian Motion, and Interacting
      Particle Systems: A Festshrift in Honor of Frank Spitzer}, eds.\
      R. Durrett and H. Kesten (Birkh\"auser, Boston, 1991).

\bibitem{G99}
       D.~J.~Grabiner, Ann. Inst. Poincare: Prob. Stat. {\bf 35}, 177 (1999). 

\bibitem{bjmkr}
      P.~L.~Krapivsky and S.~Redner, J.\ Phys.\ A {\bf 29}, 5347 (1996); 
      D.~ben-Avraham, B.~M.~Johnson, C.~A.~Monaco, P.~L.~Krapivsky,
      and S.~Redner,  J.~Phys.~A {\bf 36}, 1789 (2003). 

\bibitem{bk-mult}
      E.~Ben-Naim and P.~L.~Krapivsky,  
      J. Phys. A {\bf 43}, 495008 (2010).

\bibitem{bkl}
      E.~Ben-Naim, P.~L.~Krapivsky, and N.~W.~Lemons, 
      Phys. Rev. E {\bf 92}, 062139 (2015).
      
\bibitem{CKK}
      K.L. Chung and M. Kac, Mem. Amer. Math. Soc. {\bf 6}, 11 (1951);
      K.L. Chung and M. Kac, Ann. Math. {\bf 57}, 604 (1953);
      H. Kesten, Indiana Univ. Math. J. {\bf 12}, 391 (1963). 
      Details of the derivation of \eqref{nav:log} using the methods developed 
      in these references will be published elsewhere.

\bibitem{KLCKM}
     T.~Koren, M.~A~Lomholt, A.~V.~Chechkin, J.~Klafter, and R.~Metzler,
     Phys. Rev. Lett. {\bf 99}, 160602 (2007).
      
       

\end{thebibliography}
\end{document}